\documentstyle[12pt]{article}
\begin{document}
\vspace*{-1in}
\begin{flushright}
CERN-TH.7280/94 \\
\end{flushright}
\vskip 65pt
\begin{center}
{\Large \bf \boldmath Bounds on the Masses and Couplings of Leptoquarks
from Leptonic Partial Widths of the $Z$}\\
\vspace{8mm}
{\bf Gautam Bhattacharyya$^{\heartsuit}$, John Ellis$^{\star}$}
and {\bf K. Sridhar$^{\diamondsuit}$}\\
\vspace{10pt}
{\it Theory Division, CERN, \\ CH-1211, Geneva 23, Switzerland.}

\vspace{80pt}
{\bf ABSTRACT}
\end{center}
The vertex corrections to the leptonic partial widths of the $Z$
induced by leptoquarks that couple leptons to the top quark
are considered. We obtain
stringent bounds on the parameter space of the masses and
Yukawa couplings of these leptoquarks,
using the latest information on the $Z \rightarrow l^+l^-$ decay
widths measured at LEP. Leptoquarks coupling with electroweak
strength to top quarks are constrained to be
heavier than several hundred GeV, at 95$\%$ C.L.
As a consequence, such leptoquarks cannot make a significant
contribution to lepton asymmetries, $\tau$ polarisation asymmetries
or $A_{LR}$.

\vspace{98pt}
\noindent
\begin{flushleft}
CERN-TH.7280/94\\
May 1994\\
\vspace{11pt}
$^{\heartsuit)}$ gautam@cernvm.cern.ch\\
$^{\star)}$ johne@cernvm.cern.ch\\
$^{\diamondsuit)}$ sridhar@vxcern.cern.ch\\
\end{flushleft}

\vfill
\clearpage
\setcounter{page}{1}
\pagestyle{plain}
In the Standard Model, leptons and quarks are introduced
as independent
degrees of freedom. However, the requirement of anomaly cancellation
relates the hypercharge assignments of the quark and lepton sectors.
It is possible that this is a manifestation of a more fundamental
symmetry relating leptons and quarks. Indeed, in several
extensions of the Standard Model, such as
Grand Unified models \cite{gut},
technicolour models \cite{farhi}, and superstring-inspired $E_6$
scenaria \cite{rizzo}, there exist new boson fields that
couple leptons to quarks. Called leptoquarks, they
are SU(3)${}_c$ triplets and carry both baryon and lepton numbers.
$A\ priori$, leptoquarks could carry spin 1 or spin 0.
It is difficult to incorporate vector leptoquarks in a consistent
low-energy theory, so we focus here on scalar leptoquarks that
are electroweak doublets and couple to leptons and quarks $via$
generalised Yukawa interactions.

\vskip 10pt
If leptoquarks also coupled to quark pairs, their exchanges
would violate lepton number (L) and  baryon number (B).
However, in that case, proton stability constrains
leptoquark masses to be comparable to the Grand Unification scale
\cite{buch}.
Therefore, the leptoquarks of phenomenological interest
cannot couple to quark pairs and do not violate
$B$ and $L$.
Bounds from flavour-changing neutral currents (FCNC)
severely constrain flavour mixing in leptoquarks, so we assume
\cite{buch,pati} that they couple to only a single
generation of leptons and a single generation of quarks.
Moreover, bounds from helicity-suppressed processes
such as $\pi \rightarrow e \nu$ decay restrict leptoquark couplings
$\lambda_{\Phi}$ so severely \cite{shank} that we assume they
are chiral, i.e. each type of leptoquark couples either to
left-handed or to right-handed quarks only.
These are called left-type and right-type leptoquarks, respectively.

\vskip 10pt
Leptoquarks that do not couple to diquarks and respect these
requirements of diagonality and chirality are constrained by
searches at $e^+e^-$, $ep$, $pp$ and $\bar p p$ colliders
\cite{collider}. The LEP Collaborations exclude any
leptoquark weighing less than 45~GeV \cite{4lep}.
The D0 collaboration excludes leptoquarks with first-generation
couplings that weigh less than 133~GeV \cite{d0},
and the CDF collaboration excludes leptoquarks weighing
less than 113 GeV \cite{cdf}.
HERA experiments exclude leptoquarks that couple to electrons with
electromagnetic strength: $\lambda_{\Phi}^2=4 \pi\alpha$, and
weigh less than 145~GeV \cite{hera}.
Besides these direct limits, there
are indirect bounds on leptoquarks coming from experiments
on parity violation in atomic physics \cite{langacker, leurer}
and from searches for flavour-violating
$Z$-decays into leptons \cite{donch}.
There are also strong FCNC bounds on left-type leptoquarks,
due to the fact that CKM mixing renders impossible the diagonality
of their couplings, which are reviewed in
Refs.~\cite{leurer, campbell}.

\vskip 10pt
In this letter, we will consider only scalar leptoquarks that
transform as doublets under electroweak $SU(2)$, and couple
the top quark $t$ to any one of the three lepton generations.
This is the variety of leptoquark that is least constrained by
the above direct and indirect limits. We also assume that
its Yukawa couplings are real.
In the case of a left-type leptoquark, electroweak
gauge invariance decrees an identical
coupling to the bottom quark $b$.
We show that the LEP measurements of
the $Z \rightarrow l^+l^-$ partial widths, $\Gamma_{ll}$,
exclude such leptoquarks if they weigh less than several
hundred GeV and couple with the electroweak strength:
$\lambda_{\Phi}^2 = g_2^2 = 4\pi \alpha / \sin^2\theta_W$
\footnote{This strength appears to us as a reasonable
standard of comparison, given the large mass of the
$t$-quark and its large Yukawa coupling: $\lambda_t^2 \approx
g_2^2$ in the Standard Model.}.
The upper limits on $\lambda_{\Phi}^2$ from $\Gamma_{ll}$
are, in fact, so tight that the leptoquark contributions
to the leptonic asymmetries, $\tau$-polarization asymmetry
and the forward-backward asymmetry $A_{LR}$
must be much smaller than
the experimental errors for any value of the leptoquark mass.

\vskip 10pt
The part of the Lagrangian that describes the couplings of the
leptoquarks to the $Z$ and to the quarks and leptons is given by
\begin{equation}
\label{e1}
{\cal{L}} = {{\lambda_\Phi}{\tilde c}_{\Phi} \over s_Wc_W}
(k_1-k_2)_{\mu} \Phi^{\dagger}\Phi Z^{\mu} + {\lambda_\Phi}
 \bar l [g_L^t P_L + g_R^tP_R] t \Phi ,
\end{equation}
where ${\tilde c}_{\Phi} = t_{3\Phi}-Q_{\Phi} s^2_W$
and $\lambda_\Phi$
is the Yukawa coupling ($s_W \equiv \sin\theta_W,
c_W \equiv \cos\theta_W$).
We now consider the process
$Z \rightarrow l^+l^-$, and compute the one-loop corrections
induced by the $t-l-\Phi$ coupling. Since the leptoquark is
assumed to couple chirally, we will have to take either
($g_L^t=1, g_R^t=0)$ for a left-type leptoquark or
($g_L^t=0, g_R^t=1)$ for a right-type leptoquark.
For the left-type leptoquark, we have to
consider both $(t,\Phi)$- and $(b,\Phi)$-induced vertex corrections
(with $g^t_L = g^b_L$) \footnote{ In a basis in which the up-quark
mass matrix is diagonal, there are also
$(d,\Phi)$ and $(s,\Phi)$-contributions.
However, these are suppressed by CKM-mixing.
We have checked that their
effects are small and we have neglected them.}
while for the right-type leptoquark, the vertex correction is
only due to $(t,\Phi)$.
For the sake of simplicity we also assume that there is only one
leptoquark multiplet at a time and there is no mass splitting
within it. This assumption is justified if $m_{\Phi} \gg m_W$,
and supported by the agreement between the CDF direct measurement
of $m_t$
and the estimate based on radiative corrections, which assumes that
no other electroweak doublet has isodoublet splitting large
enough to contribute significantly to the isospin-violating
parameter $\Delta\rho$, also known as $T$ or $\epsilon_1$. We
note in passing that such a degenerate
electroweak-doublet leptoquark does not contribute to
$S$ ($\epsilon_3$) or $U (\epsilon_2)$.

\vskip 10pt
The relevant triangle and self-energy diagrams for the $Z \rightarrow
l^+l^-$ vertices are shown in Fig.~1.
Following Passarino and Veltman \cite{pasvel}, we compute the
amplitudes for the diagrams in Fig.~1 in terms of the $B$- and $C$-
functions corresponding to the two- and three-point integrals,
respectively.
In terms of the generic internal masses $m_1$ and $m_2$,
the $B$-functions are defined as
\begin{eqnarray}
\label{e2}
B_0 &\equiv & {1 \over  \pi^2} \int d^4k
{1 \over (k^2+m^2_1) \lbrace
(k-p)^2+m^2_2 \rbrace} , \nonumber \\
B_{\mu} & \equiv & {1 \over  \pi^2} \int d^4k
{k_{\mu} \over (k^2+m^2_1)
\lbrace (k-p)^2+m^2_2 \rbrace} \equiv -p_{\mu} B_1,
\end{eqnarray}
and the $C$-functions as
\begin{equation}
\label{e3}
C_0, C_{\mu}, C_{\mu\nu} \equiv {1 \over  \pi^2} \int d^4k {1,
k_{\mu}, k_{\mu\nu} \over (k^2+m^2_1)
\lbrace (k-p)^2+m^2_2 \rbrace \lbrace (k-p')^2+m^2_2 \rbrace} ,
\end{equation}
with
\begin{eqnarray}
\label{e4}
C_{\mu} &\equiv& -p_{\mu}C_{11}+q_{\mu}C_{12},
{}~~~~~~~~~~~(q = p - p')  \nonumber \\
C_{\mu\nu} &\equiv& p_{\mu}p_{\nu}C_{21}+q_{\mu}q_{\nu}C_{22}
-(p_{\mu}q_{\nu}+q_{\mu}p_{\nu})C_{23}+g_{\mu\nu}C_{24} .
\end{eqnarray}
The amplitudes
for the set of diagrams shown in Fig.~1 can be written as
\begin{equation}
\label{e5}
M_{\mu}^{(i)} = {N_c\over 16 \pi^2} {e {\lambda^2_\Phi} \over s_Wc_W}
\bar l (p^{\prime}) \gamma_{\mu} A_i l(p) ,
\end{equation}
where $i = 1,2,3$. Here $i = 1,2$ denote the contributions from the
first and the second triangle diagrams, and the contribution of
the two self-energy diagrams are jointly denoted by $i=3$.
For the sake of simplicity, we present the expressions
for the $A_i$ for the right-type leptoquark (which involve
only the top quark inside the loop):
\begin{eqnarray}
\label{e6}
A_1 &=&  \left[a_L^t m_t^2 C_0 -
a_R^t \lbrace M_Z^2 (C_{22}-C_{23}) + 2C_{24} \rbrace \right] P_L ,
 \nonumber \\
A_2 &=&-2  {\tilde c}_{\Phi} {\tilde C}_{24} P_L, \nonumber \\
A_3 &=&  a_L^l B_1 P_L.
\end{eqnarray}
In the above expressions we have taken
$N_c = 3$, and $a_L^f$ and $a_R^f$ are the tree-level
$Z$ couplings
to the left- and right-handed fermion-flavour $f$, given by
\begin{equation}
M_\mu^{tree} = {e \over{s_Wc_W}} \bar{f}(p^\prime) \gamma_\mu
(a_L^f P_L + a_R^f P_R) f(p).
\end{equation}
where
\begin{eqnarray}
a_L^f & = & t_3^f - Q_f s_W^2, \nonumber  \\
a_R^f & = & - Q_f s_W^2.
\end{eqnarray}
For a left-type leptoquark, the appropriate chirality modifications
in eq.(\ref{e6}) can be worked out trivially and we do not write them
explicitly.

\vskip 10pt
We point out that the contributions from the individual diagrams
are divergent, namely
$C_{24}$ in $A_1$, ${\tilde C}_{24}$ in $A_2$
and $B_1$ in $A_3$. But the divergence cancels when these amplitudes
are added, and we are left with a finite
correction to the partial width $Z \rightarrow l^+l^-$:
\begin{equation}
\label{e8}
\delta \Gamma_{ll} = {\alpha(M_Z) M_Z \over {3 \bar{s}_W^2
\bar{c}_W^2}}
a^l_H \delta a^l_H,
\end{equation}
where
\begin{equation}
\label{e9}
\delta a^l_H = {{\lambda^2_\Phi} \over 16\pi^2} N_c \sum_{j=1}^3 A_j.
\end{equation}
Note that we have introduced
$\bar{s}_W$ as an effective weak angle measured at the $Z$ scale,
and have put the relevant energy scale of the electromagnetic coupling
strength, $\alpha$, in the last two equations.
The index $H$ is $R$ for a left-type leptoquark and
$L$ for a right-type leptoquark.

\vskip 10pt
Leptoquark loop diagrams analogous to those considered above
{\it also contribute to the photon-electron-electron vertex}.
The diagrams
for the photon are identical to those in Fig.~1, with the $Z$ lines
replaced by photon lines.
To illustrate this point, we again take the case of the right-type
leptoquark, and substitute in
eq. (\ref{e6}) the $Z$ parameters by the photon ones; i.e.
we replace $a_L^t, a_R^t, a_L^e$ and $\tilde{c}_\Phi$
by $Q_u, Q_u, Q_e$ and $Q_\Phi$,
respectively. It is then straight-forward to check that the sum,
$\delta a_H^l$ (photon), at zero momentum transfer,
is not zero, and this has to be adjusted
against a counter term contribution of $-\delta a_H^l$ (photon),
to ensure exact charge conservation. Then,
gauge invariance fixes the corresponding counter term for the $Z$ vertex
and, hence, we get the expression of the renormalised amplitude
for the $Z$-vertex as,
\begin{equation}
\delta a_H^l (\mbox{\rm renormalised}) = \delta a_H^l + \sin^2 \theta_W
{}~\delta a_H^l (\mbox{\rm photon}).
\end{equation}
Taking this finite renormalisation into account
leads to the following modified version
of eq. (\ref{e8}), which we employ for our numerical evaluations
\begin{equation}
\label{e8mod}
\delta \Gamma_{ll} = {\alpha(M_Z) M_Z \over {3 \bar{s}_W^2
\bar{c}_W^2}}
a^l_H~\delta a^l_H (\mbox {\rm renormalised}),
\end{equation}
where $H$ is $R$ or $L$, as before, depending whether left-type
or right-type leptoquark is under investigation.

\vskip 10pt
For ease of interpretation, we present the analytic form of the
leptoquark-induced
correction in the asymptotic limit when $m_\Phi \gg M_Z$. As
before, we present explicitly only the right-type leptoquark case.
The {\it finite part} of
the sum of the $A_i$ in eq. (\ref{e6}), plus the counter term
contribution as obtained above, becomes in this limit
\begin{equation}
\sum_{j=1}^3 A_j + \sin^2\theta_W \sum_{j=1}^3 A_j (\mbox{\rm photon})
= \biggl\lbrack (a_L^t-a_R^t) \eta_2(x) +
{{M_Z^2}\over{3 m_t^2}} \left\{a_R^t \eta_1(x) +
\tilde{c}_\Phi \eta_3(x)\right\}
\biggr\rbrack,
\label{e6sim}
\end{equation}
where
($x = m_t^2/m_\Phi^2$)
and $\eta_1, \eta_2$ and $\eta_3$ are given by
\begin{eqnarray}
\eta_1(x) & = & {{-11x+18x^2-9x^3+2x^4}\over{6(1-x)^4}}
- {{x\ln x}\over{(1-x)^4}} \simeq 0~ (x \rightarrow 0),  \nonumber \\
\eta_2(x) & = & -{{x}\over{1-x}} - {{x\ln x}\over{(1-x)^2}}
\simeq 0~ (x \rightarrow 0) , \\
\eta_3(x) & = & {{2x-9x^2+18x^3-11x^4}\over{6(1-x)^4}}
+{{x^4\ln x}\over{(1-x)^4}} \simeq 0~ (x \rightarrow 0),\nonumber
\end{eqnarray}
exhibiting decoupling in the limit of large $m_\Phi$.

\vskip 10pt
To obtain limits on the leptoquark mass and the coupling parameters,
we now compare these calculations with the experimental values of the
leptonic decay widths $\Gamma_{ee}$, $\Gamma_{\mu\mu}$,
$\Gamma_{\tau\tau}$.
We parametrize ${\lambda^2_\Phi} = g_2^2 k$, where $k=1$ corresponds
to a leptoquark coupling with the electroweak strength.
In Fig.~2 we present $\delta \Gamma_{ee}$ as a function of $m_{\Phi}$,
for $k=1$.
We have evaluated the
$B$- and $C$-functions required using the code developed in
the ref.\cite{amit}, cross-checking
the results by using the standard Feynman parametrisation of the two-
and three-point functions and then integrating them
numerically. The right- and left-type leptoquark contributions
$\delta \Gamma_{ee}$ for couplings of weak $SU(2)$ strength and
$m_t=150, 165$ and 180~GeV are shown in Fig.~2
by solid, dashed and dotted lines, respectively.
Both the right- and left-type leptoquarks contribute negatively
to $\delta\Gamma_{ee}$
and are consequently constrained by the experimental lower limits.
We show in Fig.~2 the 95\% lower limits on $\delta\Gamma_{ee}$
obtained from the present experimental value
$\Gamma_{ee} = 83.96 \pm 0.22$ MeV \cite{LEP}, by subtracting
the Standard Model contribution evaluated for fixed $M_H=$ 250~GeV
and $\alpha_s(M_Z)=$~0.12 (it is quite insensitive to these
choices) and the same values $m_t=$150, 165 and 180~GeV as
previously indicated by the horizontal, solid, dashed and dotted
lines, respectively. We see that, even for $k=1$, a left-type
leptoquark up to about 680~GeV is excluded for $m_t=180$~GeV
and a right-type leptoquark weighing up to about 280~GeV.
Since the leptoquark-induced contribution $\delta\Gamma_{ee}$
is comparable to the experimental uncertainty in $\Gamma_{ee}$,
there is significant scope for increased statistics and reduced
systematic errors to place significantly stronger bounds on the
leptoquark parameter space.

\vskip 10pt
We show in Fig.~3 the constraints on leptoquarks coupling the
top quark to $e$, $\mu$ and $\tau$ in the two-parameter space
$(m_{\Phi},k)$, obtained by analogous studies of corrections
to $\Gamma_{ee}$, $\Gamma_{\mu\mu}=83.90 \pm 0.31$~MeV and
$\Gamma_{\tau\tau}=84.07 \pm 0.36$~MeV.
The maximum values of $k$ allowed
by the three leptonic partial widths for both right- and
left-type leptoquarks are shown
for the same three values of $m_t$ considered
in Fig.~2. In the case of the $t-\tau$ coupling,
the upper bound on $k$
is smaller than that for the $t-e$ coupling, in spite of the fact that
$\Gamma_{ee}$ has a smaller uncertainty than $\Gamma_{\tau\tau}$.
This is simply because the Standard Model prediction
for $Z \rightarrow \tau^+\tau^-$ happens to be
closer to the experimental 95\% C.L. lower limit.

\vskip 10pt
The leptoquark-induced corrections to the $Zl^+l^-$ couplings
also show up in the asymmetries $A_{LR}$, $A^l_{FB}$ and the
$\tau$ polarisation parameters $A_{POL}^{\tau}$ and $P_{\tau}^{FB}$.
However, the present experimental errors on these quantities are
considerably larger than the maximal leptoquark contributions
allowed by the width constraints obtained above. As an example,
we consider $A^e$, which has the same theoretical expression
as $A_{LR}$ and $P_{\tau}^{FB}$, namely
\begin{equation}
\label{aeeqn}
A^e = {{r^2 - 1}\over{r^2 + 1}}; ~~~~~~ r = {{a_L^e}\over{a_R^e}}.
\end{equation}
The leptoquark contribution to this is given by
\begin{equation}
\delta A^e = {{4r}\over{(r^2+1)^2}} \delta r,
\end{equation}
where
\begin{equation}
\delta r = {\delta a_L^e \over a_R^e} -
{\delta a_R^e \over (a_R^e)^2}a_L^e.
\end{equation}
The maximum values of $\delta r$ and hence of $\delta A^e$
(folded with the maximum allowed $k$)
allowed for the right- and left-type leptoquarks can be evaluated
quite easily using Fig.~3 and Eq.~(\ref{e9}), and are displayed
in Table 1.

\vskip 10pt
There has been considerable interest in these asymmetry and
polarisation measurements recently, stimulated in particular
by the apparent discrepancy between the recently published SLD
measurement of $A_{LR}$ and the LEP precision measurements.
It is difficult to interpret this in terms of new physics beyond the
Standard Model, particularly because $A_{LR}$ and $P_{\tau}^{FB}$
have the same theoretical expression (Eq.~\ref{aeeqn}), but differ
experimentally from each other ($0.1628 \pm 0.0077$ \cite{slac} vs.
$ 0.120 \pm 0.012$ \cite{LEP}, respectively) and lie on opposite
sides of the value expected from other measurements. Nevertheless, we
have shown in Table 1 the discrepancy between the SLD measurement
of $A_{LR}$ and the Standard Model prediction for $m_t$=150, 165,
and 180~GeV and $M_H=$ 250~GeV. We see that the apparent discrepancy
is much larger than the largest possible contribution of a
leptoquark allowed by the width analysis summarised in Fig.~3.
Therefore, a leptoquark could not explain the $A_{LR}$
measurement, even if the difference with the $P_{\tau}^{FB}$
measurement were to be resolved in its favour \footnote{We
comment in passing that the types of leptoquarks considered
here also contribute to $Z\rightarrow \bar b b$ and $\bar\nu\nu$
decays, but that these are also negligible, in view of the previous
bounds from $\Gamma_{ee,\mu\mu, \tau\tau}$.}.

\vskip 10pt
To conclude, we have placed bounds on the masses and Yukawa couplings
of $SU(2)$-doublet scalar leptoquarks with $(l,t)$ couplings
using the latest measurements
of the leptonic partial width of $Z$ at LEP. These leptoquarks
evade previous bounds \cite{shank,leurer} because of their chiral
and diagonal couplings to third-generation quarks. Moreover, as
statistics on the $Z$ peak accumulate and a better understanding of
the detectors leads to smaller systematic errors, our bounds can
be improved significantly. Further analysis on precision electroweak
constraints on these and other varieties of leptoquark is in progress
\cite{bes}.
\vskip 10pt
\noindent{\bf Acknowledgements}
\par
We thank A.~Gurtu and E.~Lisi for help and useful discussions.

\newpage
\section*{Figure captions}
\renewcommand{\labelenumi}{Fig. \arabic{enumi}}
\begin{enumerate}
\item
The one-loop Feynman diagrams contributing to
the $Z \rightarrow e^+e^-$ vertex
correction due to a leptoquark.

\item
The leptoquark-induced contribution ($\delta \Gamma_{ee}$) to
the electronic partial width of the $Z$ as a function of
$m_\Phi$, for $k=1$.
The two sets of curves correspond to the left-type (L)
and right-type (R) leptoquarks. The
solid, dashed and dotted lines correspond to $m_t =$ 150, 165 and
180 GeV, respectively.
The three horizontal lines correspond to the
allowed values of $\delta\Gamma_{ee}$ for the same choices
of $m_t$, obtained by
computing the differences between the corresponding Standard Model
predictions and the 95$\%$ C.L. lower limit obtained from
the experimental data:
$\Gamma_{ee} = 83.96 \pm 0.22$~MeV.

\item
The maximum value of $k$ obtained by comparing the left-type (L)
and the right-type (R)
leptoquark-induced contributions and the experimentally-allowed
window for new physics in leptonic partial widths shown in Fig.~2.
Curves are shown for $e, \mu$ and $\tau$
final states for our previous choices of $m_t =$ 150, 165 and 180~GeV
(solid, dashed and dotted lines, respectively).

\end{enumerate}

\newpage
\begin{table}[htbp]
\begin{center}
\caption[] {Maximum allowed contributions from left-type (L) and
right-type (R) leptoquarks to $A^e$ for $m_t =$ 150, 165 and 180 GeV.
The differences between the experimental 95$\%$ upper (lower) limits
and the Standard Model predictions are also shown.}
\bigskip
\begin{tabular}{|c|c|c|c|c|}
\hline
 & \multicolumn{2}{|c|}{Maximum leptoquark}
 & \multicolumn{2}{|c|}{Experimental} \\
 & \multicolumn{2}{|c|}{contributions}
 & \multicolumn{2}{|c|}{uncertainties} \\
\cline{2-5}
$m_t$ & $\delta A^e(L)$ & $\delta A^e(R)$ & $\delta P_{\tau}^{FB}$
(LEP) & $\delta A_{LR}$ (SLAC) \\
(GeV) &  &  &  &  \\
\hline
150 & 0.0028 & $-0.0022$ & 0.0075 & 0.0417 \\
    &        &         & ($-0.0405$) & (0.0109) \\
165 & 0.0047 & $-0.0036$ & 0.0037 & 0.0379 \\
    &        &         & ($-0.0443$) & (0.0071) \\
180 & 0.0066 & $-0.0051$ & $-0.0004$ & 0.0338 \\
    &        &         & ($-0.0484$) & (0.0030) \\
\hline
\end{tabular}
\end{center}
\end{table}

\end{document}